# *Improved Web Accessibility Evaluation of Open Learning Contents for Individuals with Learning Disabilities*

**Muhammad Ishaq**
Institute of Computer Science & Information Technology
Faculty of Management & Computer Sciences (FMCS)
The University of Agriculture, Peshawar
Peshawar-Pakistan
drmishaq@aup.edu.pk

*Web content should be accessible to normal and disabled communities on electronic devices. The Web Accessibility Initiative (WAI) has created standard guidelines called Web Content Accessibility Guidelines (WCAG). Mobile Web Best Practice (MWBP) is also proposed by WAI for accessibility of websites on desktop computers and mobile devices like smartphones, tablets, iPads, iPhones, and iPods. Educational Resources that provide free licensed learning content are used to test the WCAG. The disabled community also has equal rights to gain access to these learning materials through electronic devices. The main purpose of this research is to evaluate these selected open educational learning materials for individuals with only learning disabilities. This research provides several recommendations to improve the accessibility level of the Learning Management Systems. Future research includes developing a more accessible learning management system with minimized or no accessibility errors.*

*Disability includes physical impairments, mental disorders, lack of cognition, learning and emotional disability. Some individuals have multiple disorders. Learning disabilities are one of them. People have difficulty learning because of an unknown factor or low intelligence quotient (IQ).*

The World Wide Web (WWW) is the biggest information system or a collection of information systems. Nowadays, human life relies to a great extent on very popular web-based information. Specifically, we need a Web for searching, integration, and mining. For more improved services from the WWW, the contents (unstructured data) need to be machine readable (structured). Linked vocabulary (ontology) building is the only way to achieve machine readability of web2 contents.

Information can be easily accessed by all electronic devices like laptops, computers, smart phones, PDAs, iPhones, iPads, and many more recently developed devices. These devices access the WWW through the http protocol using any suitable browser. The goal of this work is to evaluate and make websites and applications more accessible to people with physical and learning disabilities. Physical disabilities are not limited to visual impairments, hearing impairments, blindness, and colour blindness. The extensive use of computers, laptops, smartphones, and innovative mobile application development for Android, IOS, and Windows-based mobile platforms gives birth to e-learning knowledge and research domains. Nowadays, electronic devices have become more popular and are equally important to all kinds of disabled individuals. People can access any type of information from websites everywhere at any time. Smart portable devices have new enhanced features and applications with evolving capabilities.

The WWW has become an inevitable part of our lives. The Web Accessibility Initiative (WAI) is dedicated to improving web accessibility for everyday users.Web accessibility means the mechanism and interaction of users with the WWW using any internet-enabled device. The World Wide Web Consortium (W3C) has its own guidelines and standards for developers to produce more accessible and user-friendly websites. Almost all Content Management Systems (CMS) and PHP Frameworks are capable of developing websites with improved web accessibility for normal and disabled communities. WAI of the W3C also developed guidelines called Web Content Accessibility Guidelines (WCAG) 1.0 and a more enhanced version, WCAG 2.0, for web content accessibility (Caldwell, B. et al., 2008). The WCAG consists of twelve guidelines and four principles. Each website must be perceivable, operable, understandable, and robust, and each guideline has success criteria or testable check points. They established three levels of conformance for the

Dr. Muhammad Ishaq

Dr. Muhammad Ishaq (Corresponding Author): Institute of Computer Science and Information Technology, Faculty of Management and Computer Sciences, Peshawar University of Agriculture, Pakistan. Email: ishaqafridipk@gmail.com

web page: Level A, Level AA, and Level AAA.W3C designed Mobile Web Best Practice (MWBP) for mobile web was published on July 29th, 2008 (Rabin, J. et al., 2008).It consists of some guidelines or best practices for making websites and web applications accessible from mobile devices, and these best practices in a mobile web can be tested by the mobileOK test. The test was developed by W3C for mobile web testing and follows section 508 of the Rehabilitation Act and the complete Americans with Disabilities Act (ADC) of 1973(Barnes, C. et al., 1997).

Websites consist of a lot of content, including images, text, animations, videos, and much more. The contents which are used for the purpose of learning are called the learning contents. Some learning content requires a paid subscription (Caswell, T. et al., 2008). A list of free educational sites can be accessed online through the Internet (Caswell, T. et al., 2008; Laaser, W. et al., 2017). The revolution and evolution in electronic technology make possible the access and sharing of learning content through electronic devices like laptops, computers, tablets, mobiles etc. E-learning refers to accessing learning or educational materials through electronic devices. People with Internet access can use and share learning materials from ordinary websites or specialized learning management systems (LMS) in real time.

Humans all over the world face some sort of disability. Physical impairments, mental disorders, lack of cognition, learning disability, and emotional disability are all examples of disability. Some individuals have multiple disorders. Also, disability in a person is by birth or can occur during a person's life at any age (Kavale, K. A. 2005). Learning disabilities are one of them. Major types of learning disabilities can be classified as dyslexia disability, dyscalculia disability, dysgraphia disability, language processing disability, and so on.

## Web Accessibility

to measure or estimate the ease of using the Web. Web accessibility means that the website, software, and any web-based tools that are accessed through the internet are designed and developed in such a way that they will be accessed by each person, whether he is normal or has disabilities. The web is the most important and daily used resource in many ways, like in education, government, employment, health care, business etc. So it is very important that the web is accessible to everyone in an easy and robust way. It is also important that the web is accessible for individuals with learning disabilities. The World Wide Web Consortium (W3C) has its own guidelines and standards for developers to produce more accessible and user-friendly websites. The W3C provides techniques, guidelines, supporting resources, and technical support that define and explain accessibility solutions (Chisholm, W. et al., 2001). The WAI of the W3C also develops guidelines called Web Content Accessibility Guidelines (WCAG) 1.0 and a more enhanced version, WCAG 2.0, for web content accessibility (Caldwell, B. et al., 2008). WCAG consists of twelve guidelines and four principles (Chisholm, W. et al., 2001).

### Web Accessibility evaluation tools

There are many free and paid tools for checking web accessibility. These tools are either online, browser extension based or offline software based. Some of the tools are: A-Tester by Evaluera Ltd, Automated Accessibility Testing Tools(AATT) by PayPal, Access Analytics by Level Access, Accessibility Checker by CKSource, Accessibility Checklist by Elservier, AccessLint by AccessLint, AChecker by Inclusive Design Research Center, MobileOk Checker by W3C(Leporini,.2006).

### Learning Difficulties

Learning disabilities define the problems that humans can face when they want to read any information, write something, or think about something. They failed to get the real meaning of this information, having some innate intelligence abnormalities. Learning disabilities cannot be fully fixed. It can only decrease with the right support at the right time. It can affect the education and social relationships of an individual in society. Some known and major types of learning disabilities are as follows (Devi, A., Prakash et al., 2019):

a) **Auditory Processing Disorder (APD)**

This problem is mainly related to hearing impairments, how the sounds of the words are transferred through the ear and how the brain processes these words. Learners with APD problems cannot understand even clear and loud spoken words. Also, learners can face the difficulty of recognizing the sound of the words (Devi, A., Prakash et al., 2019). Individuals with this disability have problems with following directions and with the proper order of sounds. It is an extremely tedious job for them to block competing sounds or background noise. APD individuals, for example, process thoughts slowly and misspell words with similar sounds. has difficulty focusing on verbal lectures or presentations.

b) **Dyslexia as a Disability**

Common problems encountered by patients are reading fluency, decoding, reading comprehension, recall, writing, spelling, and sometimes speech impairment. Common problems encountered by patients are reading fluency, decoding, reading comprehension, recall, writing, spelling, and sometimes speech impairment. Dyslexia also has a very bad impact on one's self-esteem and motivation.(A.Devi, Prakash et al., 2019; D. C. Geary, 2004).

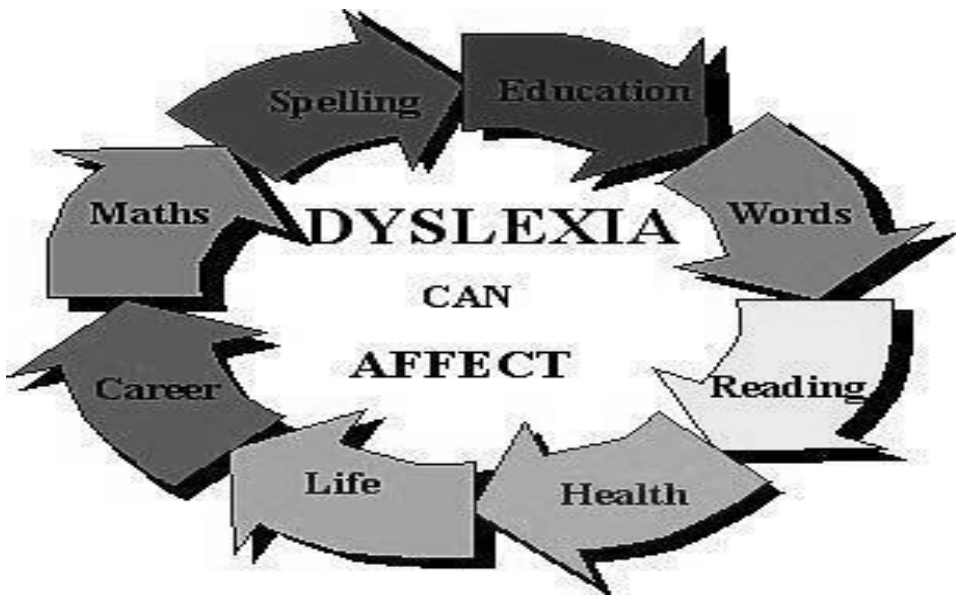

Fig. 1. Dyslexia Disability

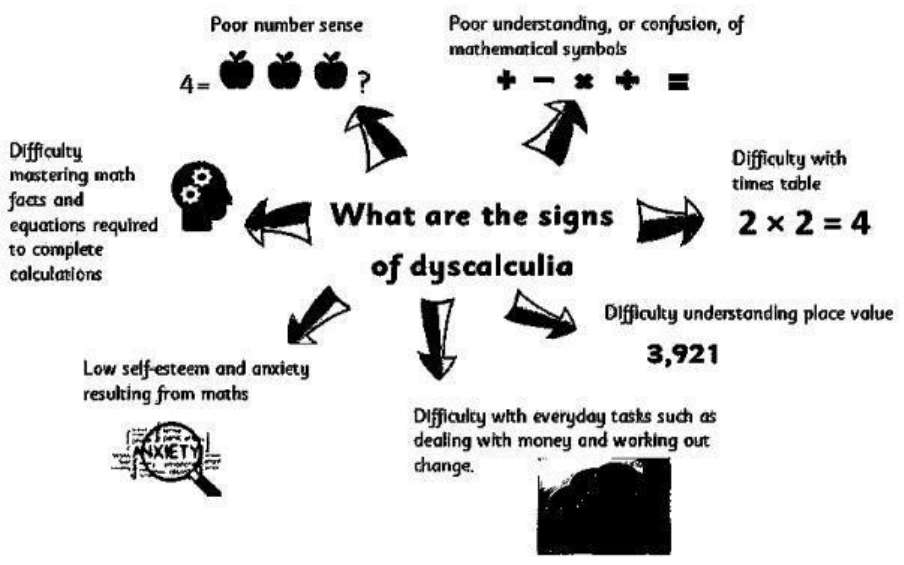

Figure 2: Disability Due to Dyscalculia

c) **Disability Due to Dyscalculia**

Learners with this disability cannot understand mathematical calculations, symbols, or numbers. People with this disability have math understanding problems at many levels. They are either reluctant to calculate or give wrong results (Devi, A., Prakash et al., 2019; Geary, D. C., 2004). Individuals are unable to remember and recognize math facts and symbols. People with this learning impairment are unable to count, memorize, and organize numbers. Individuals with dyscalculia have problems in varying ways.

d) **Dysgraphia Impairment**

People with this disability have difficulty writing words or text. The person can have difficulty with word spelling and can't recognize the space between words and cannot properly think about sentences.This disability can be minimized by supporting the individual in school, helping with individual in-home and occupational therapy(OT). This therapy can help him with arm posture and position (Jaeger, P. T. 2002).

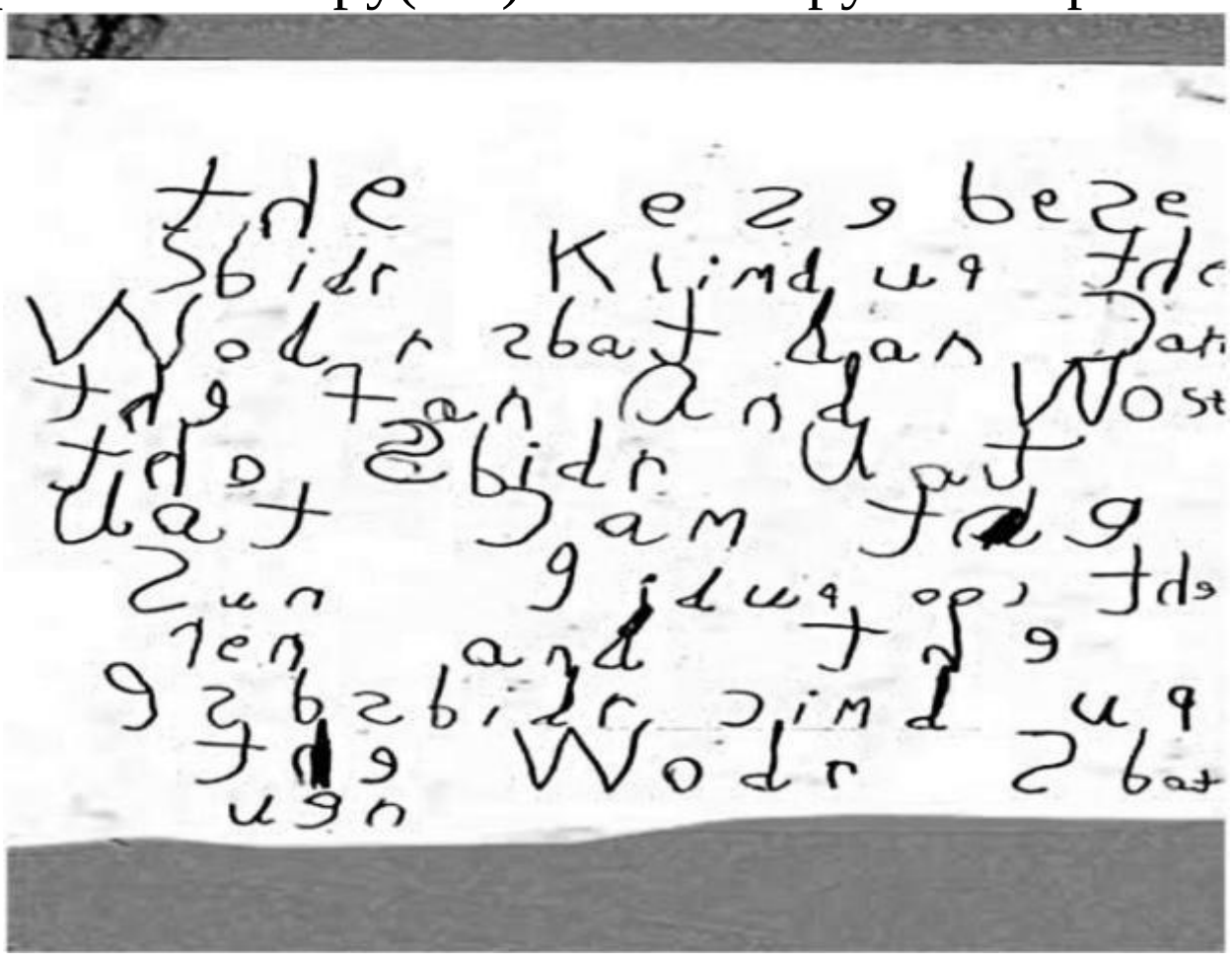

Fig. 3. Dysgraphia Disabilities

e) **Language Processing Impairment**

Individuals with language processing disabilities cannot understand the language. They face a problem when characters are combined to make a word or sentence in a language and they cannot properly understand the sounds of those words or sentences. The problem occurs only in the processing of the language (Devi, A., Prakash et al., 2019.

## Problems Addressed in This Work and Objectives of Evaluation

Web technologies are evolving day by day with innovative features. Learning through an electronic device is easy for all individuals, including the diverse disabled community. Instructors can share learning materials through Web-based learning and content management systems (LCMS) for a variety of learners with Internet-enabled devices. The goal is to give the learner secure and convenient access.Users can get on-demand or need-based knowledge from these websites using their devices.

Many online resources are available free of charge with Open Educational Resources (OER) licensing. Some learning resources are on payment and collect fees and subscription charges. Teachers or educators can share learning content on the Open Educational Resource websites like audio, video lectures, assignments, courses, syllabuses, lesson plans, textbooks, presentations, and a lot of other learning materials(Johari, et al. 2012).

This research is concerned with the evaluation of the web learning content for individuals with learning disabilities. To access the accessibility of Open Education Resources that provide online learning materials for devices and to check their accessibility for normal individuals. Also, how the different kinds of disabled people can easily find or access learning materials through their electronic devices. The main objective of this research is to

To check the accessibility of efficient open educational resources on electronic devices.To evaluate the accessibility of learning materials for individuals with learning disabilities.

## Relevant Work

A number of research papers are presented about the evaluation of web accessibility and learning disabilities for different devices.

There are several tools available to check the accessibility of websites (Leporini, et al., 2006). They believe that by using the tools, any web site can be evaluated and provide better reports for the evaluators after checking multiple guidelines for a specific web site. (Leporini, et al., 2006) designed a tool called MAGENTA, and using this tool they found web accessibility and usability. This tool checks the guidelines in the Web pages. In the case of failure or error detection, it provides a warning result (Devi, A., et al., 2019) to improve the specific area or part of the website and provide support for correcting the element. Results were generated according to the specific guidelines (Eastman et al., 2006) of W3C. Multiple pages were also evaluated by having two panel reports that were global and page-specific.

Many websites were developed for mobile devices, and their evaluation can be done automatically with the help of available tools like EvalAccess.

The work on accessibility of dynamic web applications for visually disabled users proposed a model to design an e-content system that will be easily accessible for visually impaired people. They stated that designing flexible web content, violating and removing unimportant content, making each piece of content easily accessible on dynamic web pages useful, and increasing accessibility for visually impaired users are all important.The model is implemented on dynamic websites (Kingsley, Okeye et al., 2014).

A method and approach to evaluate web accessibility for low vision users is proposed by Patricia(Acosta-Vargas et al., 2019). Through web applications, an online web tool for identifying learning disabilities is used to find a learning disability as well as recommend educational activities and works that are best suited for individuals with learning disabilities (Devi.A et al. 2019).

The use of AChecker for evaluating accessibility of e-content based on WCAG 2.0 guidelines is also common(Shah Alam et al. 2016). AChecker is used as an online free tool to evaluate web accessibility. Results of this research show that there are six main critical accessibility errors. The errors include non-text, information, link purpose, language of page, and labels. They give recommendations and instructions to improve the accessibility of e-content.

Silvia B. focuses on the accessibility evaluation of web content that contains mathematics, geometry and physics. Nowadays the use of ICT in educational institutes is increasing tremendously. Web pages that contain math, physics and geometry related contents are evaluated by Identified Authoring Tools based on WCAG 2.1 (Silvia Baldiris et al 2019).

Some Researchers design and develop a mobile based application for improving learning of dyslexic children with writing disability (Rabbia Tariq et al 2016). They develop a mobile application for android devices which helps children with writing disabilities in improving writing. Main purpose of this research is to enhance the introductory writing skill of individuals having writing difficulties.

(Mazeyanti M.A. et al. (2017) focus on developing a mobile application for those who have dyscalculia disabilities. Dyscalculia is a learning disability in which children cannot learn math easily. They developed a mobile application named Calculic Kids for improving and enhancing math learning. This Android-based application is beneficial to children with dyscalculia.

(Zainab P et al. 2015) established an e-learning system for individuals with learning disabilities. They proposed a system known as the Assistive Learning System (ALE) to improve the learning skills of individuals who have learning disabilities online. The accessibility of the web is evaluated using an automatic tool (Arrue et al. 2007). According to the valid reference of mobile web best practice, the developers can develop accessible web content since they have no expertise in mobile websites and the developers want their applications and contents. So they can be evaluated with the help of automatic tools that give them guidance while developing websites. Arrue and Vigo (Arrue et al. 2007) presented an automatic flexible evaluation tool that is EvalAccess, and this EvalAccess tool has been transformed into EvalAccess Mobile for web accessibility on mobile. This allows them to check the accessibility of web content for mobile devices. The architecture of the EvalAccess Evaluation tool consists of four modules: the Accessibility Guidelines Repository, Accessibility Guidelines Manager, Accessibility Evaluation Module, and Accessibility Report Manager (Arrue et al. 2007). This tool takes the URL of the website or web page or HTML code and checks it according to existing guidelines in Accessibility Guidelines Manager. After evaluating, Accessibility Report Manager produces a report in the form of XML which shows the warning, errors, and general warning. They tested some websites and got efficient results (Arrue et al. 2007). This tool is useful for checking the web accessibility for mobile devices and producing whether web contents meet the standard guidelines or not in the form of warnings and errors.

Evaluations of web accessibility can be tested for some specific mobile devices. How to select a tool for checking the accessibility of specific mobile devices is proposed ( Vigo Markel et al. 2008). They selected the mobileOK Basic tests created by the W3C as a basis. After the testing process, the CC/PP based profiles are generated like WURFL and UAProf files, which are in xhtml format. This mobileOK test follows the Default Delivery Context (DDC) (Vigo Markel et al. 2008). In order to check the feasibility of the tool, they selected three mobile devices and compared them with the DDC. The main objective was to select any evaluation tool that follows the mobileOK basic test, but any mobile device can be tested, not only for specific mobiles. As a result, it was concluded that the number of mobile devices has different characteristics that differ from the default delivery context. So the mobileOK basic test is useless because they consider that every mobile device follows the Default Delivery Context.

A mobile web learning platform was created by (XiaoChun. Zhou et al. 2010).As mobile is the latest technology in the world , websites should be accessed on mobile devices. That is, all pages and contents should be understandable and accessible to all communities, including normal, disabled, and elderly people. They described the importance and some problem issues regarding web mobile accessibility that were display size, text input, bandwidth, and lack of navigation (XiaoChun. Zhou et al. 2010). All web contents should adhere to the W3C's Mobile Web Best Practices for content accessibility on mobile devices (Isa, W. A. R. W. M. et al. 2016).They concluded that mobile learning platforms bring the opportunity to access learning materials on mobile devices to everyone, everywhere, at any time.

(Kalpana et al. 2012) stated that mobile devices that support internet facilities can be accessed by physically challenged people. Before their publication, it is important for developers and designers to evaluate the web contents' accessibility. After understanding the problems that are faced by persons with disabilities (PWD), they visited two well-known canters, such as Action for Ability Development and Inclusion (AADI) and another, All India Deaf and Dumb Society (AIDDS) in Delhi. They prepared a list of questions regarding PWD problems. They elicit responses from them, and using these responses, as well as their needs and demands, they created Mobione, a mobile-based acceptable toolkit designed for iPhone and iPad mobile phones.

Many websites and applications have been developed for accessibility of learning content through mobile media devices, not only for teenagers but also for children and elderly people. Many authors created various learning content management systems to provide learning materials on desktop computers and mobile devices, so mobile devices can be used to access learning content. (Judge, S. et al. 2014) described how the learning system is very efficient for young children having some disabilities using their mobile devices like smartphones, iPods, iPads, tablets, and other touch-based mobile media devices. Sharon Judge (Judge, S et al. 2014) provides a strategy of how mobile learning can give benefits to young children and how the integration of mobile devices and applications is necessary for young children who have some disabilities like visual impairments, hearing impairments, cognitive limitations, learning disabilities, color blindness, speech disabilities, photosensitivity and much more (Laaser, W. et. al 2017). Also, Sharon Judge (Judge, S et al. 2014) provides necessary information for teachers and parents which can assist them in choosing the most appropriate mobiles and applications for their children. Sharon Judge (Judge, S et al. 2015) described how the relationship between teachers with children and parents with their children should be friendly and positive. Through this, children can learn easily using new technology rather than playing with learning toys and books. A child can interact with mobile phones and apps that teach the alphabet, letter sounds, and simple words through animations. She concluded that mobile learning does not completely replace the current learning system, but if it is properly implemented, then the use of this learning system increases the experimental learning of children.

**A summary of the literature cited**

The literature review section shows that there are various efforts, developments, and contributions made by researchers to the field of web accessibility evaluation and learning disability research. The researcher works on web accessibility evaluations on different free and paid tools. Learning disabilities are discussed for the purpose of making flexible and robust web learning content for individuals with learning disabilities. Some of them develop web-based solutions and some develop mobile-based applications in order to improve the learning power of those who have learning disabilities. Therefore, this research will propose a method of web accessibility evaluation for individuals having learning disabilities.

## Accessibility Checking Mechanism

First of all, explain the procedure of the research that will be laid out. Then explain the various steps and flow of research. It explains how the research will be conducted and which steps will be followed. The free online web accessibility tools will be used to evaluate the Open Education Resources for individuals with learning disabilities. Also, explain the web accessibility evaluation for mobile devices using the MobileOk tool.

.

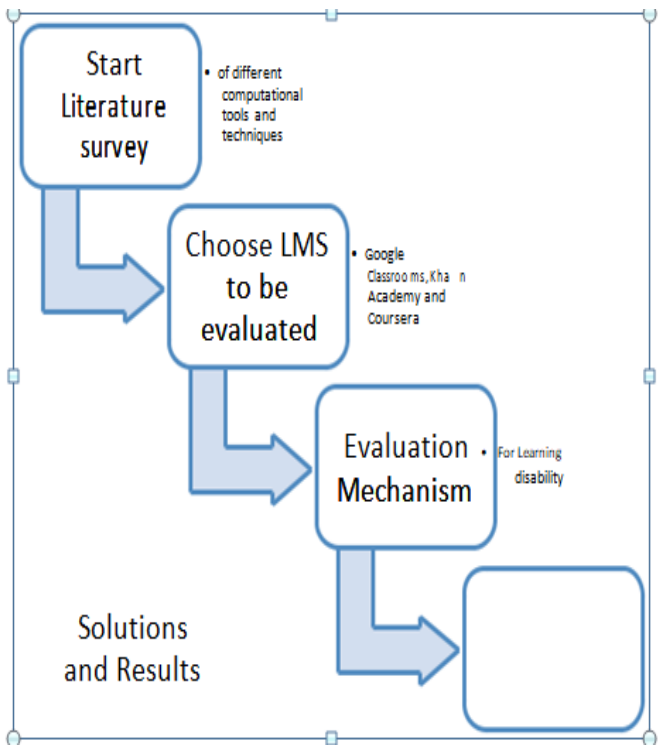

Fig 3. Research Process Flow Chart

## a) Evaluation

To evaluate the website for testing accessibility, whether the website is accessible to normal and disabled individuals or not. Evaluation can be performed with the help of tools or by expert evaluators. Many tools are available for evaluation. Some of them are online tools such as Mauve, AChecker, Magenta, Wave, and much more. Some of them are offline as software like SortSite Evaluation, HiSoftware Compliance Sheriff Web. Software tools can be paid for and some are available for free. Web browsers come with accessibility extensions like Wave and HTML validator for Chrome and Firefox. Evaluation can be done automatically or manually. With an automatic tool, whole websites are checked by entering a URL into the tool or adding an HTML file, whereas with a manual approach, specific pages or guideline check points can be evaluated manually. To assess mobile web accessibility, the W3C created a mobileOK basic test checker. The results of three famous LMS resources through AChecker are given in Table 1.1.

## b) Strategy

So, according to this research, I will adopt the manual approach. My research is about finding some open educational resources that provide learning content, and this content can be accessible or not on electronic devices for people with only learning disabilities. So I will select five open educational learning materials and evaluate them with the help of WAAT (Web Accessibility Assessment Tool), which is used for only web accessibility evaluation (Leporini, Barbara et al., 2006). This tool checks all of the WCAG 2.0 guideline checkpoints on the website for Level A priority, Level AA priority, and Level AAA priority conformance. Also, I can select specific WCAG 2.0 Success Criterions manually in this tool. So, as guidelines were designed for web accessibility and I need those guidelines which are only necessary for learning disabilities, I will select specific checkpoints and then only these checkpoints will be checked in an open educational learning content. That's why I will use the WAAT tool and select specific success criteria manually for testing the open educational resources one by one.

And for mobile web accessibility, I will select some best practices from MWBP 1.0 that will be necessary for learners with learning disabilities to access the learning content on mobile devices, and to evaluate these best practices, I will select the mobileOK basic checker. In addition, I will use other online and offline tools to evaluate the efficiency of these 05 Open Educational Learning Contents for normal accessibility and compare them to mobile web accessibility.According to the specific learning disability guidelines, I will check these guidelines into the web manually by myself and will also use the upcoming tool that is OzArt, which is also used for accessibility evaluation (Mahnegar, F. et al. 2012).

**c) Solution**

After evaluation, the result will be generated in the form of a table that consists of warnings, errors, accessibility, compatibility, usability, and some other issues in each open educational resource. It will show which web page has an issue and how much accessibility it has for learning disabilities. Which LMS is more accessible than others? This will be presented in the form of a percentage. Sample results are given in table 1.1.

## Results and Analysis

This section explains the results of the work that was done on the free web accessibility evaluation tools, namely Achecker. The results were shown in the form of tables. The first section A defines the basic preliminaries of this research. The second section is about accessibility checkers, namely Achecker. The third section, C, shows the details of WCAG 2.0. Section D contains the detailed results and discussion of this research. Section E has given the recommendation for improvement in web pages. And section F is the benefits of accessible web pages for disabled individuals.

### Preliminaries

The experiment was done on free web content or learning management systems (LMS), namely (i) Edmodo, (ii) Google Classroom, and (iii) Khan Academy. We checked the accessibility of these three LMSs using Achecker tools. Achecker is a free online web content evaluation tool based on WCAG.

Table 1.1 shows the WCAG 2.0 principles and guidelines, on the basis of which we evaluate LMS. And table 1.2 shows the results obtained through Achecker based on WCAG 2.0.

### Achecker

Achecker is a free online and open-source tool that can be used for accessibility testing of web pages based on the Web Content Accessibility Guidelines(WCAG) presented by the World Wide Web Consortium(W3C). Achecker is used to check and evaluate the accessibility of web content. There are three methods used for checking:

- By entering the URL of the web page
- Uploading the HTML complete file
- By pasting the complete code which is obtained from the web page.

Achecker can give three types of problems or errors.

- Known Problems/Errors
- Possible Issue/Error
- Potential Issue/Error

**a) Known Issue/Error:**

These are those problems which were identified by Achecker with certainty. It is necessary to fix these errors.

**b) Probable Problem/Error:**

These are those problems or errors which are identified by Achecker on the basis of probability. There is a probability of being found and not being found. and needs human confirmation. If found, then I need to fix them.

**c) Potential Issue/Error:**

Achecker is incapable of identifying issues that require human confirmation and decision.In many cases, these errors need only confirmation, whether the error is found or not. If found, then they need to modify and fix the errors on the web page. (Greg Gay and Cindy Qi Li, 2010).

## WCAG Specifications

The World Wide Web Consortium (W3C) created WCAG for evaluating web content.The LMS was evaluated on the basis of WCAG 2.0 principles and guidelines. WCAG 2.0 has four main principles that are:

PerceivableOperableUnderstandable Robust

Each principle has its own guidelines which must be satisfied by each web page for accessibility. We can check these guidelines for web pages through online accessibility evaluation tools like Achecker. Copy the link (URL) of the web page into the Achecker search field and then click the check button.

The Achecker will check the web page for accessibility based on WCAG. The detailed guidelines are listed below in the table. (Caldwell, Cooper, et al. 2008).

**Table 1.** WACG 2.0 Guidelines and principles, on the basis of which we evaluate LMS

| Principle | Guidelines | | |
|---|---|---|---|
| 1. Perceivable | 1.1 Text Alternatives | 1.1.1 | Non-Text content |
| | 1.2 Time Based Media | 1.2.3 | Audio desc / media AltObject |
| | 1.3 Adaptable | 1.3.1 | Info & Relationship: Input |
| | | 1.3.3 | Sensory Characteristics Table, Body |
| | 1.4 Distinguishable | 1.4.1 | Use of color: Body Img |
| | | 1.4.2 | Audio Control |
| | | 1.4.3 | Contrast: Link, text color |
| | | 1.4.4 | Resize Text |
| | | 1.4.5 | Images of text: img |
| | | 1.4.6 | Contrast (enhanced) |
| 2. Operable | 2.1 Keyboard Accessible | 2.1.1 | Keyboard: onmouseover |
| | | 2.1.2 | No keyboard Trap: Applet, embed |
| | 2.2 Enough Time | 2.2.2 | Pause, stop, hide |
| | 2.3 Seizures | 2.3.1 | Three Flashes : Img script |
| | 2.4 Navigable | 2.4.1 | Bypass Block |
| | | 2.4.2 | Page Titled |
| | | 2.4.4 | Link Purpose |
| | | 2.4.5 | Multiple Ways: Body |
| | | 2.4.6 | Heading & Labels |
| | | 2.4.8 | Location |
| | | 2.4.10 | Section Heading |
| 3. Understandable | 3.1 Readable | 3.1.1 | Language of page |
| | | 3.1.2 | Language of Parts |
| | | 3.1.3 | Unusual Words |
| | | 3.1.4 | Abbreviation |
| | 3.2 Predictable | 3.2.1 | On Focus |
| | | 3.2.2 | On Input: Area, Select |
| | | 3.2.3 | Consistent Navigation |
| | | 3.2.4 | Consistent Identification |
| | 3.3 Input Assistance | 3.3.1 | Error identification |
| | | 3.3.2 | Labels or Instruction |
| | | 3.3.3 | Error Suggestion |
| | | 3.3.4 | Error prevention (legal, Financial) |
| | | 3.3.6 | Error Prevention (All) |
| 4. Robust | 4.1 Compatible | 4.1.1 | Parsing: Body |
| | | 4.1.2 | Name, Role, Value |

## Results Evaluation

Table 1.2 shows the accessibility analysis based on the errors of checkpoints from the three learning management systems. The analysis details show us that there are two types of errors. The known problems or errors and any potential problems

There are a total of 6 known problems for Khan Academy, 1 for Edmodo, and 6 for Google Classroom, which were found after accessibility checking through Achecker.

Almost 287 potential errors/problems were found for Khan Academy, 34 for Edmodo, and 91 for Google Classroom.

**Table 2. Determining Errors through Achecker**

| Error Checkpoints | | | | LMS (Known Problems) | | | LMS (Potential Problems) | | |
|---|---|---|---|---|---|---|---|---|---|
| Principle & Guidelines | Success / Failure Criteria | | | Khan Academy | Edmodo | Google Classroom | Khan Academy | Edmodo | Google Classroom |
| 1. Perceivable | 1.1 | 1.1.1 | Non-Text content | 4 | 0 | 1 | 0 | 0 | 3 |
| | 1.2 | 1.2.3 | Audio desc / media AltObject | 0 | 0 | 0 | 0 | 0 | 0 |
| | 1.3 | 1.3.1 | Info & Relationship: Input | 0 | 0 | 2 | 3 | 3 | 5 |
| | | 1.3.3 | Sensory Characteristics Table, Body | 0 | 0 | 0 | 1 | 1 | 1 |
| | 1.4 | 1.4.1 | Use of color: Body Img | 0 | 0 | 0 | 35 | 8 | 15 |
| | | 1.4.2 | Audio Control | 0 | 0 | 0 | 0 | 0 | 0 |
| | | 1.4.3 | Contrast: Link, text color | 0 | 0 | 0 | 0 | 0 | 0 |
| | | 1.4.4 | Resize Text | 1 | 0 | 0 | 0 | 0 | 0 |
| | | 1.4.5 | Images of text: img | 0 | 0 | 0 | 0 | 0 | 1 |
| | | 1.4.6 | Contrast (enhanced) | 1 | 0 | 0 | 4 | 0 | 2 |
| 2. Operable | 2.1 | 2.1.1 | Keyboard: onmouseover | 0 | 0 | 0 | 30 | 8 | 10 |
| | | 2.1.2 | No keyboard Trap: Applet, embed | 0 | 0 | 0 | 0 | 0 | 0 |
| | 2.2 | 2.2.2 | Pause, stop, hide | 0 | 0 | 0 | 0 | 0 | 0 |
| | 2.3 | 2.3.1 | Three Flashes : Img script | 0 | 0 | 0 | 30 | 8 | 10 |
| | 2.4 | 2.4.1 | Bypass Block | 0 | 0 | 0 | 2 | 2 | 2 |
| | | 2.4.2 | Page Titled | 0 | 1 | 0 | 1 | 0 | 1 |
| | | 2.4.4 | Link Purpose | 0 | 0 | 0 | 172 | 0 | 14 |
| | | 2.4.5 | Multiple Ways: Body | 0 | 0 | 0 | 1 | 1 | 1 |
| | | 2.4.6 | Heading & Labels | 0 | 0 | 0 | 7 | 0 | 2 |
| | | 2.4.8 | Location | 0 | 0 | 0 | 1 | 1 | 1 |
| | | 2.4.10 | Section Heading | 0 | 0 | 0 | 1 | 1 | 1 |
| 3. Understandable | 3.1 | 3.1.1 | Language of page | 0 | 0 | 0 | 0 | 0 | 0 |
| | | 3.1.2 | Language of Parts | 0 | 0 | 0 | 1 | 1 | 1 |
| | | 3.1.3 | Unusual Words | 0 | 0 | 0 | 1 | 1 | 1 |
| | | 3.1.4 | Abbreviation | 0 | 0 | 0 | 2 | 2 | 2 |
| | 3.2 | 3.2.1 | On Focus | 0 | 0 | 0 | 0 | 0 | 0 |
| | | 3.2.2 | On Input: Area, Select | 0 | 0 | 0 | 0 | 0 | 0 |
| | | 3.2.3 | Consistent Navigation | 0 | 0 | 0 | 1 | 1 | 2 |
| | | 3.2.4 | Consistent Identification | 0 | 0 | 0 | 1 | 1 | 1 |
| | 3.3 | 3.3.1 | Error identification | 0 | 0 | 0 | 0 | 0 | 1 |
| | | 3.3.2 | Labels or Instruction | 0 | 0 | 1 | 0 | 0 | 16 |
| | | 3.3.3 | Error Suggestion | 0 | 0 | 0 | 0 | 0 | 1 |
| | | 3.3.4 | Error prevention (legal, Financial) | 0 | 0 | 0 | 0 | 0 | 1 |
| | | 3.3.6 | Error Prevention (All) | 0 | 0 | 0 | 0 | 0 | 1 |
| 4. Robust | 4.1 | 4.1.1 | Parsing: Body | 0 | 0 | 0 | 0 | 0 | 0 |

## Recommendations for Improvement

In our research, there are 19 guidelines of WCAG 2.0 which are violated by LMSs. The following gives and provides recommendations for improvement based on WCAG 2.0 principles and guidelines, which were defined by the World Wide Web Consortium (W3C).

**a. Nontextual Content**

This standard defines and describes the non-text content presented to users in web content. These non-text elements must have text alternatives. It will cause confusion and errors if an image is in an empty link or if an image has very long alternative text in the web content alt attributes. This guideline is the basic requirement in web content design, so it is important for designers to solve this problem. The length of the alternative text must be more than or among 100 characters and must indicate the content that is not available to the end user. To solve this problem, suitable alternative text must be added that explains the content of the image in case the link is broken or not available. Web developers and designers must provide alternative text that is descriptive and expressed clearly. (Caldwell, Cooper, et al. 2008).

**b. Guideline:Information & Relationships.**

To solve this problem, web developers must ensure that the label for form controls is displayed by using the <label> tag. If the label was not found, then provide the associated labels. Another thing is that a form with a required field must be labeled either with an asterisk character (*) or a text label. (Caldwell, Cooper, et al. 2008).

**c. Guideline: Sensory Characteristics (Table, Body):**

Instructions provided for understanding and operating content do not provide for understanding and operating content and do not solely rely on sensory characteristics of components such as shape, color, size, visual location, etc. The intent of these success criteria is to ensure that all users can access instruction for using the contenteven when they cannot perceive the shape, size, or use of information. The success criteria require that additional information be provided to clarify anything that is dependent on this kind of information. (Caldwell, Cooper, et al. 2008).

**d. Guideline: Color Scheme: Body, IMG:**

Color is a vital thing in web content design. Therefore, the use of color for information or data in such a way that each and every user can use and understand it Some users have difficulty perceiving colors, like people with color blindness. As a result, color must be used in an efficient manner so that the contrast between image and text color is such that everyone can easily differentiate between image and text. (Caldwell, Cooper, et al. 2008).

**e. Guideline: Text Illustration:**

The text, which is used for visual presentation, can be used in the form of text rather than in the form of an image of the text. If a web developer or designer can use text to get the same visual effect, then he must use text instead of images of text. To achieve the same result and effect, use CSS. (Caldwell, Cooper, et al. 2008).

**f. Guideline: Contrast (Enhanced)**

The success of this guideline is to provide enough contrast between text, images, and their backgrounds. Small text needs higher contrast, and large text has broader characters and needs low contrast to read easily. Ensure that a contrast ratio of 7:1 exists between text, images, and background. Provide a style switcher to switch to high contrast. (Caldwell, Cooper, et al. 2008).

**g. Guideline: Keyboard (OnMouseOver):**

The web content must be operable through a keyboard. This guideline must be followed by the developer. This helps people with low vision who have difficulties tracking mouse pointers on screen. (Caldwell, Cooper, et al. 2008).

**h. Guidelines: Three Flashes: Img Script:**

The developer and designer need to design web pages in such a way that they do not contain anything that flashes or blinks more than three times in a one-second period. Blinking is a source of distraction.Keep the flashing area as small as possible.(Caldwell, Cooper, et al. 2008).

**i. Guideline: Bypass Block:**

The success criteria of this guideline is to allow people who navigate sequentially through content more direct access to the primary content of the web page. Add a link on the top of each page that goes directly to the main content. (Caldwell, Cooper, et al. 2008).

**j. Guideline: Page Title:**

Web pages must have a title that describes the purpose of the page's content. A title is used to quickly identify what the information in the content is about. Give descriptive titles to web pages. (Caldwell, Cooper, et al. 2008).

**k. Guideline: Link Purpose (In Content):**

The purpose of each link can be determined by the link text alone or by the link text together. The intent of this guideline's success criterion is to help users understand the purpose of each link so they can decide whether they want to follow this link or not. Providing text that describes the purpose of the link. Providing texts that describe the purpose of links for anchor elements (Caldwell, Cooper, et al. 2008).

**l. Guideline: Multiple Ways: Body:**

The intent of this guideline is to make it possible for users to locate content in a manner that best meets their needs. Providing an opportunity to navigate sites in more than one manner can help people find information quickly.

Make a table of contents.

Present links to navigate to related pages.

Please provide a list of links to all other web pages. (Caldwell, Cooper, et al. 2008).

**m. Guideline: Heading and Labels:**

The heading and labels must be in such a way that they describe the purpose of the content. Descriptive labels help users identify specific components within the content. (Caldwell, Cooper, et al. 2008).

**n. Guideline: Location:**

The intent of this guideline is to provide a way for the user to orient himself within a set of web pages and find related information. It will help people with short attention spans. It provides a site map to navigate easily. Using the navigation bar to indicate the current location (Caldwell, Cooper, et al. 2008).

**o. Guidelines: Section Heading:**

It is used to organize the content in better ways. The success of this guideline is to provide a heading for each section of a web page. People with learning disabilities will be able to use the heading to understand the overall organization of the page content more easily. Using headings to organize a page. Provide a heading element at the start of each section of content.(Caldwell, Cooper, et al. 2008).

**p. Guideline: Language of Parts:**

This guideline's success criteria is to ensure that content is written in multiple languages. Both assistive technologies and conventional user agents can render text more accurately if the language of each passage of text is identified. Screen readers can use the pronunciation rules of the language of text. This is important when switching between languages. People with learning disabilities who use text-to-speech software.

(Caldwell, Cooper, et al. 2008).

**q. Guideline: Unusual words:**

Certain disabilities make it difficult to understand non-literal words and specialized words or usage. This guideline success criterion will help people with language and learning disabilities.

who have trouble decoding wordsas well as those who struggle to understand words and phrasesWe can eliminate this problem as we do not use such words or, if we do, then link these words with their definition and use description list. (Caldwell, Cooper, et al. 2008).

**r. Guideline: Abbreviation:**

An abbreviation is the short form of a word, phrase, or name. The success criteria of this guideline is to ensure that users can easily get its expanded form. must connect abbreviation and definitionProviding an explanation and expansion of the abbreviation (Caldwell, Cooper, et al. 2008).

**s. Guideline: Consistent Navigation:**

All navigation elements are the same on every web page and occur in the same order each time they are repeated, unless a change is made by the users. It uses recognition rather than recall. Help those with learning and visual impairments! (Caldwell, Cooper, et al. 2008).

**t. Guideline: Consistent Identification:**

Components that have the same functionality within a set of web pages are identified consistently. People who learn functionality on one page on a site can find the desired functionality on other pages. Use labels, names, and text alternatives consistently for content that has the same functionality. (Caldwell, Cooper, et al. 2008).

**u. Guideline: Error Identification:**

The success of this guideline is to ensure that users are aware that an error has occurred and can determine what is wrong. The error message should be as specific as possible. Providing information about input errors It can help people with learning disabilities who have difficulties understanding the meaning represented by icons and other visual cues. (Caldwell, Cooper, et al. 2008).

**v. Guideline: Labels & Instruction:**

Labels or instructions are provided when content requires input. e.g. label for input fields. Name for the input field, etc.If there is no visible label, then there is a need to provide an associated label. Add a descriptive title attribute to the form control. (Caldwell, Cooper, et al. 2008).

**w. Guideline: Error Suggestion:**

The intent of this success criterion is to ensure that users receive appropriate suggestions for correction of an input error if possible. Providing information about how to correct input errors allows users who have learning disabilities to fill out the form successfully. Provide a text description to identify required fields that are not completed. must use validation to form controls. (Caldwell, Cooper, et al. 2008).

x. **Guideline: Error Prevention**:

Users must prevent errors on web pages that are used for legal or financial transactions. Submissions are reversible. A mechanism must be available for reviewing, confirming, and correcting information before finalizing the submission. People with disabilities may be more likely to make mistakes. People with reading disabilities may transpose numbers and letters. (Caldwell, Cooper, et al. 2008).

**y. Guideline: Error Prevention (All)**

All errors must be prevented, and possible prevention techniques must be used. Data entered by the user is checked for input errors, and the user is provided an opportunity to correct these errors. Submissions are reversible. Provide a mechanism for reviewing, confirming, and correcting information before finalizing the submission. (Caldwell, Cooper, et al. 2008).

## Conclusion And Future Work

This research proposed the accessibility evaluation of an online free learning management system for individuals with learning disabilities. A lot of relevant research work is referred to. The accessibility evaluation of open educational learning for learning disabled individuals is our novel contribution. Evaluation of the accessibility of open educational learning content is based upon the Web Content Accessibility Guidelines (WCAG 2.0).

This research evaluates the problem and errors which were found on the web pages of open educational learning resources. Detailed explanation of issues and reasonable recommendations on how to handle or solve these problems. Overall, the research gives good results and a better way to eliminate and decrease problems in the web pages of free learning management systems.

In this pandemic, online learning management systems play a vital role in all academic and research activities throughout the world.

This research proposed an online method to check the accessibility of web pages through online free accessibility evaluation tool Achecker. In this research, most work has been done for evaluation of open educational content. This research uses the free online tool Achecker for checking the accessibility of web pages of learning management systems and evaluating the problem/errors on the basis of Web Content Accessibility Guideline(WCAG 2.0). Correct identification of issues may lead us to develop online learning resources that enhance accessibility for normal and disable individuals. Online learning is of utmost importance in general and vital in this pandemic hit situation. Nearly all Academic and research oriented organizations need individuals that can identify potential issues in existing Electronic learning resources. There is a calamity of experts that can design and develop learning materials with improved accessibility for all types of learners. The use of virtual reality and computer vision for improved content development and design is a future work perspective in this area. An enhanced virtual reality based learning system may solve many issues for individuals with learning disabilities.

## DECLARATION OF CONFLICTING INTERESTS

The author declared no potential conflicts of interest with respect to the research, authorship and/or publication of this article.

## FUNDING

This is self sponsored research and there is no finanncial support for the research, authorship and/or publication of this article.